\documentstyle[12pt]{article}
\newcommand{\dd}{\mbox{\rm d}}

\newcommand{\nnn}{\noindent}
\newcommand{\p}{\partial}
\newcommand{\oo}{\over}

\newcommand{\be}{\begin{equation}}
\newcommand{\ee}{\end{equation}}
\newcommand{\lbl}{\label}
\newcommand{\bi}{\bibitem}

\newcommand{\vs}{\vspace}

\begin{document}
\baselineskip .7cm

\thispagestyle{empty}

\ 

\vs{1.5cm}

\begin{center}
{\Large \bf Pseudo Euclidean-Signature Harmonic Oscillator, Quantum
Field Theory and Vanishing Cosmological Constant}
\vs{6mm}

Matej Pav\v si\v c\footnote{Email: MATEJ.PAVSIC@IJS.SI}

Jo\v zef Stefan Institute, Jamova 39, SI-1000 Ljubljana, Slovenia

\vs{1.5cm}

ABSTRACT
\end{center}

The harmonic oscillator in pseudo euclidean space is studied. A
straightforward procedure reveals that although such a system may
have negative energy, it is stable. In the quantized theory 
the vacuum state has to be suitably defined and then the zero-point
energy corresponding to a positive-signature component is canceled by
the one corresponding to a negative-signature component. This principle
is then applied to a system of scalar fields. The metric in the space of
fields is assumed to have signature $(+ + + ... - - -)$ and it is shown
that the vacuum energy, and consequently the cosmological constant, are then
exactly zero. The theory also predicts the existence of stable, negative
energy field excitations (the so called "exotic matter") which are sources
of repulsive gravitational fields, necessary for construction of the time
machines and Alcubierre's hyperfast warp drive. 

\vs{1cm}

\nnn PACS: 0320, 0370, 0490 

\nnn KEY WORDS: Harmonic oscillator, pseudo euclidean space, negative
energy density, vacuum energy, cosmological constant

\vs{9mm}

\nnn Phone: +386 61 177-3780 ; Fax: +386 61 219-385  or 273-677

\newpage

{\bf 1. Introduction}

\vs{.2cm}

Quantum field theory is a very successful theory and yet it is not free
of some unresolved problems. One of the major obstacles to future progress
in our understanding of the relation between quantum field theory and
gravity is the problem of the cosmological constant \cite{1}. Since a quantum
field is an infinite set of harmonic oscillators, each having a
non vanishing zero-point energy, a field as a whole has infinite
vacuum energy density . The latter can be considered,
when neglecting the gravitational field, just as an additive constant
with no influence on dynamics. However, the situation changes
dramatically if one takes into account the gravitational field
which feels the absolute energy density. As a consequence the
infinite (or more realistically, given by Planck scale cutoff)
cosmological constant is predicted which is in drastical contradiction
with its observed small value. No generally accepted solution to
this problem has been found so far.

In the present paper I study the system of uncoupled harmonic oscillators
in a space with arbitrary metric signature. One interesting consequence
of such a model is vanishing zero-point energy of the system when
the number of positive and negative signature coordinates is the
same and so there is no cosmological constant problem. As an example
I first solve a system of two oscillators in the space with signature
$(+ -)$ by using a straightforward, though surprisingly unexploited
approach, based on the fact that energy is here just a quadratic
form consisting of positive and negative terms. Both positive and negative
energy component are on the same footing, and the difference in sign does
not show up, until we consider gravitation. In the quantized theory
the vacuum state can be defined straightforwardly and the zero-point
energies cancel out. If the action is written in a covariant notation,
one has to be careful of how to define the vacuum state. Usually
it is required that energy be positive, and then, in the absence
of a cutoff, the formalism contains an infinite vacuum energy and
negative norm states. In the proposed formalism energy is not necessarily
positive, negative energy states are also stable and there is no negative
norm states.

The formalism is then applied to the case of scalar fields. The extension of
the spinor-Maxwell field is also discussed. Vacuum
energy density is zero and the cosmological constant vanishes.
However, the theory contains the negative energy fields which
couple to the gravitational field in a different way than the usual,
positive energy, fields: the sign of coupling is reversed. This is
the prize to be paid, if one wants to get small cosmological
constant in a straightforward way. One can consider this as a prediction
of the theory to be tested by suitably designed experiments.
Usually, however, the argument is just the opposite to the one
proposed in this paper and classical matter is required to satisfy
certain (essentially positive) energy conditions  \cite{1a} which can
only be violated by quantum field theory.

\vs{1cm}

{\bf 2. The 2-dimensional pseudo euclidean harmonic oscillator}

\vs{.2cm}

Instead of the usual harmonic oscillator in 2-dimensional space
let us consider the one given by the Lagrangian
\be
     L = {1\oo 2} ({\dot x}^2 - {\dot y}^2) - {\omega^2 \oo 2}
     (x^2 - y^2)
\lbl{1}
\ee

\nnn The corresponding equations of motion are
\be
       {\ddot x} + \omega^2 x = 0 \> , \qquad {\ddot y} + \omega^2 y = 0
\lbl{2}
\ee

\nnn Note that in spite of the minus sign in front of the $y$-terms
in the Lagrangian (\ref{1}), $x$ and $y$ components satisfy
the same type equations of motion.

The canonical momenta conjugate to $x$,$y$ are
\be
    p_x = {{\p L}\oo {\p \dot x}} = \dot x \> , \qquad
       p_y =  {{\p L}\oo {\p \dot y}} = - \dot y
\lbl{5}   
\ee

\nnn The Hamiltonian is
\be
       H = p_x \dot x + p_y \dot y - L = {1\oo 2} (p_x^2 - p_y^2)
       + {{\omega^2}\oo 2} (x^2 - y^2)
\lbl{5a}
\ee

\nnn We see immediately that the energy so defined may have positive
or negative values, depending on initial conditions. Even if the system
happens to have negative energy, it is stable, since the particle
moves in a closed curve around the point (0,0). Motion of the harmonic
oscillator based on the Lagrangian (\ref{1}) does not differ from the
one of the usual harmonic oscillator. The difference occurs when one
considers the gravitational fields around the two systems.

The Hamiltonian equations of motion are
\be
       \dot x = {{\p H}\oo {\p p_x}} =  \lbrace x, H \rbrace = p_x
       \; , \quad
       \dot y = {{\p H}\oo {\p p_y}} = \lbrace y, H \rbrace  = - p_y
\lbl{7}
\ee

\be
      {\dot p}_x = - {{\p H}\oo {\p x}} = \lbrace p_x,H \rbrace =
      - \omega^2 x \; , \quad
      {\dot p}_y = - {{\p H}\oo {\p y}} = \lbrace p_y,H \rbrace =
      \omega^2 y
\lbl{9}
\ee

\nnn where the basic Poisson brackets are $\lbrace x, p_x \rbrace = 1$
and $\lbrace y, p_y \rbrace = 1$.

Quantizing our system we have
\be
       \lbrack x, p_x \rbrack = i \; \; \quad \quad [y, p_y] = i
\lbl{10}
\ee

\nnn Introducing the non hermitian operators according to
\be
    c_x = {1\oo \sqrt{2}} \left ( \sqrt{\omega} + {i\oo {\sqrt{\omega}}} p_x
     \right ) \> , \qquad c_x^{\dagger} = 
     {1\oo \sqrt{2}} \left ( \sqrt{\omega} - 
     {i\oo {\sqrt{\omega}}} p_x \right )
\lbl{11}
\ee        
         
\be
    c_y = {1\oo \sqrt{2}} 
    \left ( \sqrt{\omega} + {i\oo {\sqrt{\omega}}} p_y \right ) \> , \qquad
    c_y^{\dagger} = {1\oo \sqrt{2}} \left ( \sqrt{\omega} - 
    {i\oo {\sqrt{\omega}}} p_y \right )
\lbl{12}
\ee        

\nnn we have
\be
      H = {\omega \oo 2} (c_x^{\dagger} c_x + c_x c_x^\dagger -
          c_y^{\dagger} c_y - c_y c_y^{\dagger})
\lbl{13}
\ee

\nnn From the commutation relations (\ref{10}) we obtain
\be
     [c_x, c_x^{\dagger}] = 1 \> , \qquad [c_y, c_y^{\dagger}] = 1
\lbl{14}
\ee

\nnn and the normal ordered Hamiltonian then becomes
\be
      H= \omega (c_x^{\dagger} c_x - c_y^{\dagger} c_y)
\lbl{15}
\ee

\nnn The vacuum state is defined as
\be
      c_x|0 \rangle = 0 \> , \qquad c_y |0 \rangle = 0
\lbl{16}
\ee

\nnn The eigenvalues of $H$ are
\be
      E = \omega (n_x - n_y)
\lbl{16a}
\ee

\nnn where $n_x$ and $n_y$ are eigenvalues of the operators
$c_x^{\dagger} c_x$ and $c_y^{\dagger} c_y$, respectively.

The zero-point energies belonging to the $x$ and $y$
components cancel out! {\it Our 2-dimensional pseudo harmonic
oscillator has vanishing zero-point energy}!. This is a result we
obtain when applying the standard Hamilton procedure to the
Lagrangian (\ref{1}).

In the $(x,y)$ representation the vacuum state $\langle x,y|0 \rangle
\equiv \psi_0 (x,y)$ satisfies
\be
      {1\oo \sqrt{2}} \left ( \sqrt{\omega} \, x + {1\oo \sqrt{\omega}} 
      {\p \oo {\p x}} \psi_0 (x,y) \right ) = 0
\; , \; \; \; 
     {1\oo \sqrt{2}} \left ( \sqrt{\omega} \, y + {1\oo \sqrt{\omega}} 
      {\p \oo {\p y}} \psi_0 (x,y) \right ) = 0
\lbl{18}
\ee
                                                    
\nnn which comes straightforwardly from (\ref{16}). A solution which
is in agreement with the probability interpretation,
\be
     \psi_0 = {{2 \pi} \oo \omega} {\rm exp}[- {\omega \oo 2} (x^2 + y^2)]
\lbl{19}
\ee

\nnn is normalized according to $\int \psi_0^2 \, {\dd} x \,{\dd} y = 1$.

We see that our particle is localized around the origin. The excited
states obtained by applying $c_x^{\dagger}$, $c_y^{\dagger}$ on the
vacuum state are also localized. This is in agreement with the fact
that also according to the classical equations of motion (\ref{2}),
the particle is localized in the vicinity of the origin.
All states $|\psi \rangle$ have positive norm. For instance,
$\langle 0| c c^{\dagger}|0 \rangle = \langle 0|[c,c^{\dagger}]|0 \rangle
= \langle 0|0 \rangle = \int \psi_0^2 {\dd} x \, {\dd} y = 1$.

\vs{1cm}

{\bf 3. Harmonic oscillator in $d$-dimensional pseudo euclidean space}

\vs{2mm}

Extending (\ref{1}) to arbitrary dimension it is convenient to use the
compact (covariant) index notation
\be
     L = {1 \oo 2} {\dot x}^{\mu} {\dot x}_{\mu} - {{\omega^2} \oo 2}
     x^{\mu} x_{\mu}
\lbl{20}
\ee

\nnn where for arbitrary vector $A^{\mu}$ the quadratic form is
$A^{\mu} A_{\mu} \equiv \eta_{\mu \nu} A^{\mu} A^{\nu}$. The metric
tensor $\eta_{\mu \nu}$ has signature $(+ + + ... - - - ...)$. The
Hamiltonian is
\be
      H= {1\oo 2} p^{\mu} p_{\mu} + {{\omega^2} \oo 2} x^{\mu} x_{\mu}
\lbl{21}
\ee

\nnn Conventionally one introduces
\be
       a^{\mu} = {1 \oo \sqrt{2}} \left ( \sqrt{\omega} \, x^{\mu} + {i \oo 
       \sqrt{\omega}} \, p^{\mu} \right )
\> , \qquad
       {a^{\mu}}^{\dagger} = {1 \oo \sqrt{2}} \left ( \sqrt{\omega} x^{\mu} - 
       {i \oo \sqrt{\omega}} p^{\mu} \right )
\lbl{23}
\ee
                  
\nnn In terms of $a^{\mu}$, $a^{\mu \dagger}$ the Hamiltonian reads
\be
      H = {\omega \oo 2} (a^{\mu \dagger} a_{\mu} + a_{\mu} a^{\mu \dagger})
\lbl{24}
\ee

\nnn Upon quantization we have
\be  [x^{\mu}, p_{\nu}] = i {\delta^{\mu}}_{\nu} \qquad {\rm or}
    \qquad [x^{\mu}, p^{\nu}] = i \eta^{\mu \nu}
\lbl{25}
\ee

\nnn and
\be
    [a^{\mu}, a_{\nu}^{\dagger}] = {\delta^{\mu}}_{\nu} \qquad {\rm or} 
    \qquad  [a^{\mu}, a^{\nu \dagger}] = \eta^{\mu \nu}
\lbl{25a}
\ee

We shall now discuss two possible definitions of vacuum state. The first
possibility is the one that is usually assumed, while the second 
possibility is the one I am proposing in this paper.

{\it Possibility I.} Vacuum state can be defined according to
\be
        a^{\mu} |0 \rangle = 0
\lbl{26}
\ee

\nnn and the Hamiltonian, normal ordered with respect to the vacuum
definition (\ref{26}), after using (\ref{25a}) becomes
 \be
       H = \omega \left ( a^{\mu \dagger} a_{\mu} + {d \oo 2}
       \right ) \; ,  \qquad d = \eta^{\mu \nu} \eta_{\mu \nu}
\lbl{27}
\ee

\nnn Its eigenvalues are all positive and there is the non vanishing
zero-point energy $\omega d/2$. In the $x$ representation the vacuum
state is
\be
  \psi_0 = \left ({{2 \pi} \oo \omega} \right )^{d/2} {\rm exp}[-{\omega \oo 2}
    x^{\mu} x_{\mu}]
\lbl{28}
\ee

\nnn It is a solution of the Schr\" odinger equation
$-(1/2) \p^{\mu} \p_{\mu} \psi_0 + (\omega^2/2) x^{\mu} x_{\mu} \psi_0
= E_0 \psi_0$ with positive $E_0 = \omega ({1\oo 2} + {1\oo 2} + ....)$.
The state $\psi_o$ as well as excited states can not be normalized to 1.
Actually, there exist negative norm states. For instance, if $\eta^{33}
= -1$, then $\langle 0|a^3 a^{3 \dagger}|0 \rangle = \langle 0|
[a^3, a^{3 \dagger}]|0 \rangle = - \langle 0|0 \rangle$.

{\it Possibility II.} Let us split $a^{\mu} = (a^{\alpha}, 
a^{\bar \alpha})$, where indices
$\alpha$, ${\bar \alpha}$ refer to the components with positive and
negative signature, respectively, and define vacuum according to\footnote{
Equivalently, one can define annihilation and creation operators in terms
of $x^{\mu}$ and the canonically conjugate momentum $p_{\mu} = 
\eta_{\mu \nu} p^{\nu}$ according to $c^{\mu} = (1/\sqrt{2}) (\sqrt{
\omega} x^{\mu} + (i/\sqrt{\omega}) p_{\mu})$ and $c^{\mu \dagger} =
(1/\sqrt{2}) (\sqrt{ \omega} x^{\mu} - (i/\sqrt{\omega}) p_{\mu})$,
satisfying $[c^{\mu}, c^{\nu \dagger}] = \delta^{\mu \nu}$. Vacuum is
then defined as $c^{\mu} |0 \rangle = 0$. This is just the higher
dimensional generalization of $c_x$, $c_y$ (eq.(\ref{11}),(\ref{12})
and the vacuum definition (\ref{16}). }
\be
    a^{\alpha} |0 \rangle = 0  \; \qquad \qquad a^{{\bar \alpha} \dagger}
    |0 \rangle = 0
\lbl{29}
\ee

\nnn Using (\ref{25a}) we obtain the normal ordered Hamiltonian with
respect to the vacuum definition (\ref{29})
\be
      H = \omega \left ( a^{\alpha \dagger} a_{\alpha} + {r \oo 2} +
           a_{\bar \alpha} a^{{\bar \alpha} \dagger} - {s \oo 2} \right )
\lbl{30}
\ee

\nnn where ${\delta_{\alpha}}^{\alpha} = r$ and ${{\delta}_{\bar \alpha}}^
{\bar \alpha} = s$. If the number of positive and negative signature
components is the same, i.e., $r = s$, then the Hamiltonian (\ref{30})
has vanishing zero-point energy:
\be
H = \omega (a^{\alpha \dagger} a_{\alpha} + 
           a_{\bar \alpha} a^{{\bar \alpha} \dagger})
\lbl{31}
\ee

\nnn Its eigenvalues are positive or negative, depending on which component
(positive or negative signature) are excited. In $x$-representation
the vacuum state (\ref{29}) is
\be
     \psi_0 = {\left ( {{2 \pi} \oo \omega} \right ) }^{d/2} {\rm exp}[-{\omega
     \oo 2} \delta_{\mu \nu} x^{\mu} x^{\nu}]
\lbl{32}
\ee

\nnn where the Kronecker symbol $\delta_{\mu \nu}$ has values +1 or 0.
It is a solution of the Schr\" odinger equation
$-(1/2) \p^{\mu} \p_{\mu} \psi_0 + (\omega^2/2) x^{\mu} x_{\mu} \psi_0
= E_0 \psi_0$ with $E_0 = \omega ({1\oo 2} + {1\oo 2} + .... - {1\oo 2} -
{1\oo 2}- ...)$. One can also easily verify that there is no negative
norm states.

Comparing {\it Possibility I} with {\it Possibility II} we observe that the
former has positive energy vacuum invariant under pseudo euclidean
rotations, while the latter has the vacuum invariant under euclidean
rotations and having vanishing energy (when $r=s$). In other words,
we have either (i) non vanishing energy and pseudo euclidean invariance or
(ii) vanishing energy and euclidean invariance of the vacuum state.
In the case (ii) the vacuum state $\psi_0$ changes under the pseudo
euclidean rotations, but its energy remains zero.

The invariance group of our Hamiltonian (\ref{21}) and the
corresponding Schr\" odinger equation consists of
pseudo-rotations. Though a solution to the Schr\" odinger
equation changes under a pseudo-rotation, the theory is 
covariant under the pseudo-rotations in the sense that the
set of all possible solutions does not change under the pseudo-rotations.
Namely, the solution $\psi_0 (x')$ of the Schr\" odinger equation
$-(1/2) {\p'}^{\mu} {\p'}_{\mu} \psi_0(x') + 
(\omega^2/2) x'^{\mu} x'_{\mu} \psi_0(x') = 0$ in a pseudo-rotated
frame  $S'$ is $\psi_0 (x') = (2\pi/\omega)^{d/2} {\rm exp} [-(\omega/2)
\delta_{\mu \nu} x'^{\mu} x'^{\nu}]$. If observed from the frame
$S$ the latter solution reads $\psi'_0 (x) =
(2\pi/\omega)^{d/2} {\rm exp} [-(\omega/2)\delta_{\mu \nu} {L^\mu}_\rho
{L^{\nu}}_\sigma]$, where $x'^\mu = {L^\mu}_\rho x^\rho$. One finds
that $\psi'_0 (x)$ as well as $\psi_0(x)$ (eq.(\ref{32}) are solutions
of the Schr\" odinger equation in $S$ and they both have the same
vanishing energy. In general, in a given reference frame we have thus
a degeneracy of solutions with the same energy. This is so also in the
case of excited states.

In principle it seem more naturally to adopt {\it Possibility II},
because classically energy of our harmonic oscillator is nothing but
a quadratic form $E = (1/2) (p^\mu p_\mu + \omega^2 x^\mu x_\mu)$ which
in the case of pseudo euclidean-signature metric can be positive,
negative or zero.

\vs{1cm}

{\bf 4. A system of scalar fields}

\vs{2mm}

Suppose we have a system of two scalar fields described by the
action\footnote
{Here, for the sake of demonstration, I am using the formalism
of the conventional field theory, though in my opinion a better
formalism involves an invariant evolution parameter \cite{2}.}
\be
    I = {1\oo 2} \int {\dd}^4 x \, (\p_{\mu} \phi_1 \, \p^\mu \phi_1 - 
    m_1^2 - \p_\mu \phi_2 \, \p^\mu \phi_2 + m^2 \phi_2^2)
\lbl{36}
\ee

\nnn This action differs from the usual action for a charged field
in the sign of the $\phi_2$ term. It is a field generalization of
our action for the point-particle harmonic oscillator in 2-dimensional
pseudo euclidean space.

The canonical momenta are
\be
        \pi_1 = {\dot \phi}_1 \; , \qquad \pi_2 = - {\dot \phi}_2
\lbl{37}
\ee

\nnn satisfying
\be
   [\phi_1({\bf x} ), \pi_1({\bf x}')] = i \delta^3 ({\bf x} - {\bf x} ')
   \> , \qquad
   [\phi_2({\bf x}), \pi_2({\bf x}')] = i \delta^3 ({\bf x} - {\bf x}')
\lbl{39}
\ee
                 
\nnn The Hamiltonian is
\be
    H = {1\oo 2} \int {\dd}^3 {\bf x} \, ({\bf \pi}_1^2 + m^2 \phi_1^2 -
    \p_i \phi_1 \, \p^i \phi_1 - {\bf \pi}_2^2 - m^2 \phi_2^2 +
    \p_i \phi_2 \p^i \phi_2)
\lbl{40}
\ee

\nnn We use the spacetime metric with signature (+ - - -) so that
$-\p_i \phi_1 \, \p^i \phi_1 = ({\bf \nabla} \phi)^2$, $i = 1,2,3$. Using the
expansion ($\omega_{\bf k} = (m^2 + {\bf k} ^2)^{1/2}$)
\be
    \phi_1 = \int {{{\dd}^3 {\bf k}}\oo {(2 \pi)^3}} \, 
    {1\oo {2 \omega_{\bf k}}}
    \left (c_1 ({\bf k} ) e^{- i k x} + c_1^{\dagger} ({\bf k} ) e^{i k x}
    \right )
\lbl{41}
\ee
\be
    \phi_2 = \int {{{\dd}^3 {\bf k}} \oo {(2 \pi)^3}} \,
    {1\oo {2 \omega_{\bf k}}}
    \left (c_2 ({\bf k} ) e^{- i k x} + c_2^{\dagger} ({\bf k} ) e^{i k x}
    \right )
\lbl{42}
\ee

\nnn we obtain
\be
    H = {1\oo 2} \int {{\dd}^3 {\bf k} \oo {(2 \pi)^3}} \, 
    {\omega_{\bf k} \oo 2 \omega_{\bf k}} \left ( c_1^{\dagger}({\bf k} ) 
     c_1({\bf k}) + c_1 ({\bf k}) c_1^{\dagger} ({\bf k}) -
     c_2^{\dagger}({\bf k} ) c_2({\bf k}) -
     c_2 ({\bf k}) c_2^{\dagger} ({\bf k}) \right )
\lbl{43}
\ee

The commutation relations are
\be
     [c_1 ({\bf k}), c_1^{\dagger}({\bf k}'] = (2 \pi)^3 2 \omega_{\bf k}
     \, \delta^3 ({\bf k} - {\bf k}')
\lbl{44}
\ee
\be
     [c_2 ({\bf k}), c_2^{\dagger}({\bf k}'] = (2 \pi)^3 2 \omega_{\bf k}
     \, \delta^3 ({\bf k} - {\bf k}')
\lbl{45}
\ee

\nnn The Hamiltonian can be written in the form
\be
    H = {1\oo 2} \int {{\dd}^3 {\bf k} \oo {(2 \pi)^3}} \, 
    {\omega_{\bf k} \oo 2 \omega_{\bf k}} \left ( c_1^{\dagger}({\bf k} ) 
     c_1({\bf k})  - c_2^{\dagger}({\bf k} ) c_2({\bf k}) \right )
\lbl{47}
\ee

\nnn If we define vacuum according to
\be
    c_1({\bf k}) |0 \rangle = 0 \; , \qquad c_2({\bf k}) |0 \rangle = 0
\lbl{48}
\ee

\nnn then the Hamiltonian (\ref{47}) contains the creation operators on the left
and has no zero-point energy. However, it is not positive definite: it
may have positive or negative eigenvalues. But, as it is obvious from our
analysis of the harmonic oscillator (\ref{1}), negative energy states in
our formalism are not automatically unstable; they can be as stable as
positive energy states.

Extension of the action (\ref{36}) to arbitrary number os fields $\phi^a$ is
straightforward. Let us now include also the gravitational field $g_{\mu \nu}$.
The action is then
\be
    I = {1\oo 2} \int {\dd}^4 x \, \sqrt{-g} \,
    (g^{\mu \nu} \p_\mu \phi^a \, \p_\nu \phi^b
    \, \gamma_{ab} - m^2 \, \phi^a \phi^b \gamma_{ab} + {1 \oo {16 \pi G}} \, R )
\lbl{49}
\ee

\nnn where $\gamma_{ab}$ is the metric tensor in the space of $\phi^a$.
Variation of (\ref{49}) with respect to $g^{\mu \nu}$ gives the Einstein
equations
\be
     R_{\mu \nu} - {1\oo 2} g_{\mu \nu} R = - 8 \pi G \, T_{\mu \nu}
\lbl{50}
\ee

\nnn where the stress-energy tensor is
\be
     T_{\mu \nu} = {2 \oo {\sqrt{-g}}} \,  {{\p {\cal L}} \oo {\p g^{\mu \nu}}}
     = \left [\p_{\mu} \phi^a \, \p_{\nu} \phi^b - {1\oo 2} \, g_{\mu \nu} \, 
     (g^{\rho \sigma} \p_{\rho } \phi^a \, \p_{\sigma} \phi^b 
     - m^2 \phi^a \phi^b ) \right ]\gamma_{ab}
\lbl{51}
\ee

\nnn If $\gamma_{ab}$ has signature $(+ + + ... - - -)$ with the same number of 
plus and minus signs, then the vacuum contribution to $T_{\mu \nu}$ cancel out,
so that the expectation value $\langle T_{\mu \nu} \rangle$ remains finite.
In particular we have
\be
     T_{00} = {1\oo 2} ({\dot \phi}^a {\dot \phi}^b - \p_i 
     \phi^a \, \p^i \phi^b + m^2 \phi^a \phi^b ) \gamma_{ab}
\lbl{52}
\ee

\nnn which is just the Hamiltonian $H$ of eq.(\ref{40}) generalized to
an arbitrary number of fields $\phi^a$.

An analogous procedure as before could be done for other types of fields
such as charged scalar, spinor and gauge fields.
The notorious cosmological constant problem does not arise in our
model, since vacuum expectation value $\langle 0| T_{\mu \nu}|0 \rangle = 0$.
We could reason the other way around: since experiments clearly show that the
cosmological constant is small, this indicates (especially in the
absence of any other acceptable explanation) that to every field there
corresponds a companion field with opposite signature of the metric
eigenvalue in the space of fields. The companion field need not be excited
- and thus observed - at all. Its mere existence is sufficient to be
manifested in the vacuum energy.

However, there is the prize to be paid. If negative signature fields are
excited, then $\langle T_{00} \rangle$ can be negative which implies repulsive
gravitational field around such a source. Such a prediction of the theory
could be considered as an annoyance on the one hand, or a virtue on the
other hand. If the latter point of view is taken, then we have here the so
called exotic matter with negative energy density which is necessary for
construction of the stable wormholes with the time machine properties \cite{3}.
Also the Alcubierre warp drive \cite{4} which enables faster than light
motion with respect to a distant observer requires negative energy
density matter.

As an example let me show how the above procedure works for spinor
and gauge fields. Neglecting gravitation the action is
\be
    I = \int {\dd}^4 x\, \left [ i {\bar \psi}^a \gamma^{\mu}
    (\p_{\mu} \psi^b + i e 
    {{A_{\mu}}^b}_c \psi^c) - m {\bar \psi}^a \psi^b +{1 \oo {16 \pi}}
    {F_{\mu \nu}}^{ac} {{F^{\mu \nu}}_c}^b \right ] \gamma_{ab}
\label{53}
\ee    

\nnn where ${F_{\mu \nu}}^{ab} = \p_{\mu} {A_{\nu}}^{ab} - 
\p_{\nu} {A_{\mu}}^{ab} - ({A_{\mu}}^{ac} {A_{\nu}}^{db} - 
({A_{\nu}}^{ac} {A_{\mu}}^{db}) \gamma_{cd}$. This action is invariant
under local rotations $\psi'^a = {U^a}_b \phi^b$, ${\bar \psi}'^a = {U^a}_b
\psi^b$, ${{A'_{\mu}}^a}_b = {{U^*}^c}_b {{A_{\mu}}^d}_c {U^a}_d +
i {{U^*}^c}_b \p_{\mu} {U^a}_c$ which are generalization of the usual
$SU(N)$ transformations to the case of the metric $\gamma_{ab} =
{\rm diag} (1,1,..., -1, -1)$.

In the case when there are two spinor fields $\psi_1$, $\psi_2$ and
$\gamma_{ab} = {\rm diag} (1,-1)$ the equations of motion derived from
(\ref{53}) admit a solution $\psi_2 =0$, $A^{12}=A^{21}=A^{22}=0$.
In the quantum field theory such a solution can be interpreted that when
the fermions of the type $\psi_2$ (the companion or negative signature fields)
are not excited (not present) also the gauge fields $A^{\mu 12}$, $A^{\mu 21}$,
$A^{\mu 22}$ are not excited (not present). What remains are just the ordinary
$\psi_1 \equiv \psi$ fermion quanta and $U(1)$ gauge field $A^{\mu 11}
\equiv A^{\mu}$ quanta. The usual spinor-Maxwell electrodynamics is just
a special solution to the more general system given by (\ref{53}).

Although having vanishing vacuum energy such a model is consistent with
the well known experimentally observed effects which are manifestations
of vacuum energy. Namely, the companion particles $\psi_2$ are expected
to be present in the earth material in small amounts at most, because
otherwise the gravitational field around the Earth would be repulsive.
So, when considering vacuum effects, there remain only (or predominantly)
the interactions between the fermions $\psi_1$ and the virtual photons
$A^{\mu 11}$. For instance, in the case of the Casimir effect \cite{5}
the fermions $\psi_1$ in the two conducting plates interact with the
virtual photons $A^{\mu 11}$ in the vacuum and hence impose the boundary
conditions on the vacuum modes of $A^{\mu 11}$ in the presence of the plates.
As a result have a net force between the plates, just like in the usual theory.

When gravitation is not taken into account, the fields within the doublet
$(\psi_1, \, \psi_2)$ as described by the action (\ref{53}) are not easily
distinguishable, since they have the same mass, charge and spin. They mutually
interact only through the mixed coupling terms in the action (\ref{53})
and unless the effects of this mixed coupling are specifically measured,
the two fields can be misidentified as a single field.
Its double character could manifest itself straightforwardly in the presence
of gravitational field to which the members of a doublet couple with the
opposite sign. In order to detect such doublets (or perhaps multiplets)
of fields, one has to perform suitable experiments. Description of such
experiments is beyond the scope of this paper which only aims to bring
attention to such a possibility. Here I only mention that difficulties
and discrepancies in measuring precise value of the gravitational constant
might have roots in negative energy matter. The latter would affect
the measured value of the effective gravitational constant, but would
leave the equivalence principle untouched.

\vs{1cm}

{\bf 5. Conlcusion}

\vs{2mm}

The problem of the cosmological constant is one of the toughest problems
in theoretical physics. Its resolution would open the door to further
understanding of the relation between quantum theory and general relativity.
Since all more conventional approaches seems to have been more or less
exploited without unambiguous success, the time is right for a more drastic
novel approach. Such is the one which relies on the properties of the harmonic
oscillator in a pseudo euclidean space. This can be applied to the field
theory where the fields behave as components of a harmonic oscillator. If the
space of fields has the metric with signature $( + + + ... - - -)$ then the
vacuum energy can be zero in the case when the number of plus and minus signs
is the same. As a consequence, expectation value of the stress-energy tensor,
the source of gravitational field, is finite, and there is no cosmological
constant problem. However, the stress-energy tensor can be negative in
certain circumstances and the matter then acquires exotic properties
which are desirable for certain very important theoretical constructions,
like the time-machines \cite{3} or faster-than-light warp drive \cite{4}.
Negative energy matter, with repulsive gravitational field, is considered here
as a prediction of the theory. On the contrary, in a more conventional approach
just the opposite point of view is taken. It is argued that, since for all
known forms of matter gravitation is attractive, certain energy conditions
(weak, strong and dominant) must be satisfied \cite{1a}. But my point of view,
advocated in this paper, is that existence of negative energy matter is
necessary in order to keep cosmological constant small (or zero). Besides
that, such exotic matter, if indeed present in the Universe, should
manifest itself in various gravitation related phenomena. Actually, we cannot
claim to posses a complete knowledge and understanding of all those
phenomena, especially when some of them are still waiting for a generally
accepted explanation.  

The theory of the pseudo euclidean-signature harmonic oscillator is possibly
important also for strings. Since it eliminates the zero-point energy,
it presumably eliminates also the need for the critical dimension. We may
thus expect to obtain a consistent string theory in an arbitrary
even dimensional spacetime with suitable signature. Several exciting
new possibilities of research are thus opened.

\vs{1cm}

\centerline{\bf Acknowledgement}

This work was supported by the Slovenian Ministry of Science and
Technology under Contract J1-7455-0106-96.

\end{document}